\documentclass[aps,pra,showpacs,groupedaddress,twocolumn,amsmath,amssymb]{revtex4}

\usepackage[american]{babel}
\usepackage{graphicx}
\usepackage{times}
\usepackage{subfigure}
\usepackage{color}

\begin{document}
%\title{On the nonclassicality of phase-diffused squeezed light}
\title{Experimental test of nonclassicality criteria}

\author{T. Kiesel, W. Vogel}

\affiliation{Arbeitsgruppe Quantenoptik, Institut f\"ur Physik, Universit\"at  Rostock, D-18051 Rostock,
Germany}

\author{B. Hage, J. DiGuglielmo, A. Samblowski, and R. Schnabel}

\affiliation{
Institut f\"{u}r Gravitationsphysik, Leibniz Universit\"{a}t Hannover
 and Max-Planck-Institut f\"{u}r Gravitationsphysik (Albert-Einstein-Institute),\\
Callinstrasse 38, 30167 Hannover, Germany
}
\begin{abstract}
We experimentally examine the nonclassical character of a class of non-Gaussian states known as
phase-diffused squeezed states. These states may show no squeezing effect at all, and therefore provide an interesting example to test nonclassicality criteria. The characteristic function of the Glauber-Sudarshan representation ($P$ function) proves to be a powerful tool to detect nonclassicality. Using this criterion we find that phase-diffused squeezed states are always nonclassical, even if the squeezing effect vanishes. Testing other criteria of nonclassicality based on higher-order squeezing and the positive semidefinitness of special matrices of normally ordered moments, it is found that these criteria fail to reveal the nonclassicality for some of the prepared phase-diffused squeezed states.
\end{abstract}


\maketitle

\section{Introduction}

The definition of nonclassicality of a quantum state of the harmonic
oscillator is closely connected to the coherent states. These are the
eigenstates of the annihilation operator,
$\hat a\left|\alpha\right> = \alpha \left|\alpha\right>$, where the complex number $\alpha$ defines
the amplitude and phase of the field \cite{CoherentStates}.
%Coherent states can be regarded as the quantum analog of a classical harmonic oscillator. DIESER SATZ ERSCHEINT MIR NICHT NOTWENDIG ZU SEIN.
Glauber and Sudarshan showed that the density operator of an
arbitrary optical quantum state can be formally written as a statistical
mixture of coherent states,
\begin{equation}
        \hat \rho = \int d^2\alpha\, P(\alpha)
        \left|\alpha\right>\left<\alpha\right|,
\end{equation}
where the Glauber-Sudarshan representation $P(\alpha)$
plays the role of the probability distribution of coherent states
\cite{SudarshanP,GlauberP}. However, in quantum optics $P(\alpha)$
often violates the properties of a probability density. Hence, a state
is referred to as nonclassical if its $P$ function does not exhibit the
properties of a classical probability density \cite{DefNonclassicality}.

Only recently, nonclassicality of experimentally generated states has been
demonstrated by means of this definition \cite{SPATS}. In many cases, however, the $P$ functions of nonclassical states are highly singular, such that they cannot be reconstructed from the measured experimental data.
This is the case for squeezed states, having a quadrature variance of
less than the quadrature variance of the vacuum state. For instance, the $P$ function
of a squeezed vacuum state with quadrature variances $V_x$ and $V_p$
(we assume that $V_x < 1$, where unity represents the normalized vacuum noise) may be formally written as
 \begin{equation}
        P_{\rm sv}(\alpha) =
        e^{-\frac{V_x-V_p}{8}\left(\frac{\partial^2}{\partial
        \alpha^2} + \frac{\partial^2}{\partial \alpha^{*2}}  -
        2 \frac{V_x+V_p-2}{V_x-V_p}\frac{\partial}{\partial
        \alpha}\frac{\partial}{\partial \alpha^*}\right)}\delta(\alpha).
\end{equation}
This quantity cannot be understood as a well-behaved function.

Phase-diffused squeezed states define an interesting class of states with a, in general, not accessible
$P$ function. In very recent experiments these states were used to demonstrate \emph{purification}
and \emph{distillation} for continuous variable quantum information
protocols \cite{FranzenSQZpuri,HageSQZpuri}.  Phase diffused squeezed
states are a mixture of squeezed (vacuum) states with a stochastically
distributed phase. They are related to a realistic decoherence process and
may be produced from pure squeezed states in a phase noisy transmission
channel. They reveal a non-Gaussian noise distribution, have a positive
Wigner function, and, for strong phase noise, may show no squeezing
effect at the level of second moments of the quadrature operators.

In this paper, we use phase-diffused squeezed states in order to
experimentally test nonclassicality criteria for the case where the
$P$ function cannot directly be reconstructed from the homodyne detector quadrature
data.  First, we concentrate
on the \emph{characteristic function} of the $P$ function, which
is always well-behaved, and investigate the criterion proposed in
\cite{VogelCFCriterion}. Second, we examine \emph{moments} of the quadrature
operator and search for higher-order squeezing \cite{HongMandel}. Third,
we check a hierarchy of criteria based on \emph{normally ordered moments}, as
suggested in \cite{AgarwalMatrices}. We find that the characteristic function
of the $P$ function outperforms the
other criteria of nonclassicality.

Let us consider a statistical mixture of squeezed states, each described
by a Wigner function \cite{Fiurasek}
\begin{equation}
        W_{\rm sv}(x,p;\varphi) = \frac{1}{2\pi\sqrt{V_xV_p}}
        \exp\left\{-\frac{x_\varphi^2}{2V_x}-\frac{p_\varphi^2}{2V_p}\right\},
\label{eq:Wigner:squeezed}
\end{equation}
where $x_\varphi = x\cos(\varphi) + p \sin(\varphi)$
and $p_\varphi = -x\sin(\varphi) + p \cos(\varphi)$ are the quadrature
variables, rotated around an angle $\varphi$, and $V_x, V_p$ are the
variances of both quadratures $x_\varphi, p_\varphi$, satisfying the Heisenberg uncertainty
relation $V_x V_p \geq 1$.
 Let $p(\varphi)$ denote the statistical
distribution of the phase fluctuations, then the Wigner function of the
mixed state reads
\begin{equation}
        W(x,p) = \int p(\varphi)
        W_{\rm sv}(x,p;\varphi)\,d\varphi. \label{eq:mixed:state:Wigner}
\end{equation}
 In our examination of nonclassicality, the characteristic function
 $\Phi(\beta)$ of the $P$ function plays a decisive role. It is connected
 to the Wigner function via Fourier transform
\begin{equation}
        \Phi(\beta) = e^{|\beta|^2/2}\int W(x,p) e^{i(x\,{\rm Im}\beta -
        p\,{\rm Re}\beta)}dx\,dp .
\end{equation}
For a squeezed state, as defined by Eq.~(\ref{eq:Wigner:squeezed}), we find
\begin{equation}
        \Phi_{\rm sv}(\beta;\varphi) = e^{\frac{|\beta|^2}{2} \left[1
        - V_x \cos^2(\arg(\beta) - \varphi) - V_p \sin^2(\arg(\beta) -
        \varphi) \right]}.\label{eq:charfunc:squeezed}
\end{equation}
The characteristic function for the mixed state is given
in close analogy to Eq.~(\ref{eq:mixed:state:Wigner}),
 \begin{equation}
        \Phi(\beta) = \int p(\varphi) \Phi_{\rm
        sv}(\beta;\varphi)\,d\varphi \label{eq:mixed:state:CF}.
\end{equation}

\begin{table}
        \begin{tabular}{c || c | c | c | c | c}
                        &\multicolumn{5}{c}{$V_x = 0.36,\ V_p = 5.28$}\\\hline
                $\sigma / {}^\circ$     &       $0.0$   &       $6.3$   &       $12.6$  &       $22.2$  &       $\infty$\\
                $V_{\rm eff}$           &       $0.36$  &       $0.42$  &       $0.59$  &       $1.00$  &       $2.82$
        \end{tabular}
        \caption{Parameter of the examined states.}
        \label{tab:states}
\end{table}

In our experiment we generated phase-diffused squeezed vacuum states with
varying strengths of the phase noise.  The phase noise was chosen to be distributed
according to a zero mean Gaussian and could therefore be completely
characterized by the standard deviation. A summary of states generated
is given in Table~\ref{tab:states}. The undisturbed squeezed vacuum states
had quadrature variances $V_x = 0.36$ and $V_p = 5.28$. For the strongest phase noise we used a flat distribution with a width of 720$^{\circ}$, which is labeled with $\sigma = \infty$  in Table~\ref{tab:states}. We also
listed the minimum quadrature variance of each state, \begin{equation}
        V_{\rm eff} = \frac{V_x + V_p}{2} - \frac{V_p - V_x}{2}
        e^{-2\sigma^2},
\end{equation}
 to show that the states with $\sigma = 6.3^\circ$ and
$\sigma = 12.6^\circ$ are still squeezed, but the squeezing vanishes at
$\sigma = 22.2^\circ$. Hence, one cannot decide about the nonclassicality
of the last two states by examination of the quadrature variance.

%\emph{theoretical proof that states are nonclassical?}
\section{Experimental set-up}
The squeezed states were generated by a degenerate optical parametric
amplifier (OPA). The OPA consisted of a type-I non-critically phase
matched second order nonlinear crystal (7\% Mg:LiNbO$_3$) inside a
standing wave optical resonator with a line width of 25$\,$MHz. The OPA process was continuously pumped
by 50$\,$mW of second harmonic light yielding a classical power amplification factor of six. Both the length (resonance frequency)
of the resonator as well as the orientation of the squeezing ellipse
were stably controlled by electronic servo loops.
 With this setup we directly measured a minimal
{squeezed} variance of -4.5$\,$dB and an {anti-squeezed}
variance of +7.2$\,$dB with respect to the unity vacuum variance. From
these measurements we inferred an overall efficiency
of 75\% and an initial squeezing factor of -8.2$\,$dB.

\begin{figure}[h]
\includegraphics[width=\linewidth]{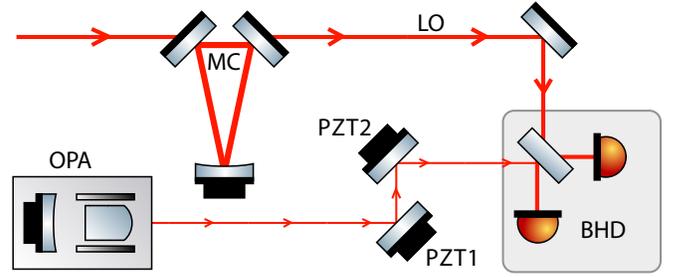}
\caption{Simplified sketch of the experimental setup. MC: spatial mode cleaner, LO: local oscillator, OPA: squeezed light source, BHD: balanced homodyne detector, PZT: piezo-electrically actuated mirror. PZT1 was used to control the average phase and PZT2 applied the phase noise.}
\label{fig:setup}
\end{figure}

The squeezed field propagated in free space from the OPA passing
high-reflection mirrors, two of which were moved by piezo-electric
transducers (PZT). One (PZT1, Fig.~\ref{fig:setup}) was used to
control the average phase of the squeezed field. The other (PZT2,
Fig.~\ref{fig:setup}) was driven by a quasi-random voltage to apply the
phase diffusion. This voltage was generated by a high quality PC sound
card connected to an appropriate amplifier. The sound card played back a
previously generated sound file which was carefully designed to meet the
desired shape of its frequency spectrum and its histogram. The former covered the \emph{flat} part of the frequency response of the PZT but the frequency band of any control loop, the latter
was chosen to be Gaussian for the partial phase diffusion and had to be
absolutely flat in the totally randomized case.

Balanced homodyne detection (BHD) was used to measure the quadrature
amplitude of the phase-diffused squeezed field. The visibility of the
squeezed beam and the spatially filtered (MC, Fig.~\ref{fig:setup}) local
oscillator was 98.9\% and was limited by OPA crystal inhomogeneities. The
average quadrature phase of the BHD was servo loop controlled except for
the total phase randomization where no mean phase exists.
 The signals
of the two individual BHD-photodetectors were electronically mixed down
at 7$\,$MHz and low pass filtered with a bandwidth of 400$\,$kHz to
address a modulation mode showing good squeezing and a high dark noise
clearance of the order of 20$\,$dB. The resulting signals were fed into
a PC based data acquisition system and sampled with one million samples
per second and 14 bit resolution.
 For a more detailed description of
the main parts of the setup we refer to \cite{HageSQZpuri,FranzenSQZpuri}.

\section{Nonclassicality in terms of the characteristic function}
\subsection{Experimental demonstration}

First, let us consider a sufficient criterion proposed in
\cite{VogelCFCriterion}: A state is nonclassical if the characteristic
function $\Phi(\beta)$ of the $P$ function exceeds the characteristic
function of {the} vacuum at some point $\beta$:
 \begin{equation}
        \exists \beta:\quad |\Phi(\beta)| > |\Phi_{\rm vac}| \equiv
        1.\label{eq:criterion:CF}
\end{equation}
Note that this condition represents the lowest order of a hierarchy of conditions which completely characterize the nonclassicality~\cite{VogelCFHierarchy}.
The function $\Phi(\beta)$ can be obtained by~\cite{vowe-book}
 \begin{equation}
        \Phi(\beta) = \left<:\hat D(\beta):\right> =
        \left<e^{i|\beta|\hat x(\pi/2-\arg(\beta))}\right>
        e^{|\beta|^2/2}, \label{eq:charfunc:P}
\end{equation}
 where $\hat D(\beta)$ is the displacement operator. Since
we only consider a single quadrature, we may neg\-lect the arguments of $\beta$
and $\hat{x}$. The expectation value on the right hand side of
Eq.~(\ref{eq:charfunc:P}) represents
the characteristic function of the quadrature. It can be estimated from
the sample of $N$ measured quadrature values $\left\{x_j\right\}_{j=1}^N$
via (cf.~\cite{lvovskyShapiro})
\begin{equation}
        \left<e^{i|\beta|\hat x}\right> \approx
        \frac{1}{N}\sum_{j=1}^{N}e^{i|\beta|x_j}. \label{eq:charfunc:x}
\end{equation}
Inserting Eq.~(\ref{eq:charfunc:x})
into~(\ref{eq:charfunc:P}), we obtain an estimation $\overline{\Phi}(\beta)$
of ${\Phi}(\beta)$.
The variance of this quantity can be estimated as
\begin{equation}
        \sigma^2\left\{\overline{\Phi}(\beta)\right\}  =
        \frac{1}{N}\left[e^{|\beta|^2}-\left|\overline{\Phi}(\beta)\right|^2\right].
		\label{eq:Sample:CF:Standard:Deviation}
\end{equation}

\begin{figure}
\includegraphics[width=\columnwidth]{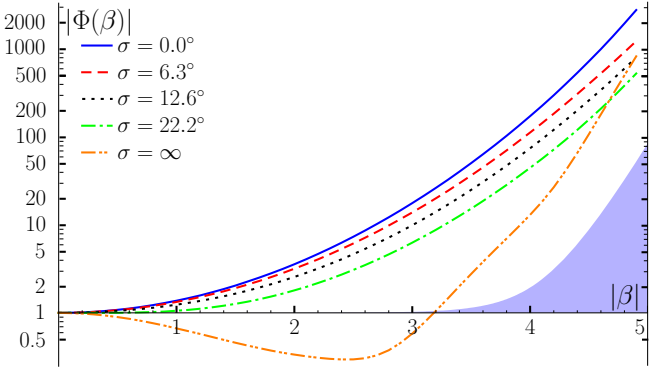}
\caption{Characteristic functions of different phase-diffused squeezed vacuum states. The shaded area corresponds to one standard deviation, it is added to the nonclassical boundary $\Phi_{\rm vac} \equiv 1$. Take
note of the logarithmic scale.} \label{fig:CF}
 \end{figure}

For each state we have recorded $N=10^7$ data points. The resulting characteristic functions are shown in
Fig.~\ref{fig:CF}. We only concentrate on the quadrature where the variance
of the state is minimum. The shaded area corresponds to the magnitude
of one standard deviation, it is added to the nonclassicality border
$|\Phi_{\rm vac}(\beta)| \equiv 1$. In order to demonstrate that a
state satisfies the nonclassicality criterion, Eq.~(\ref{eq:criterion:CF}),
with a significance of $s$ standard deviations, we have to check if the
characteristic function {satisfies} the inequality
\begin{equation}
        |\overline{\Phi}(\beta)| >  1 + s
        \sigma\left\{\overline{\Phi}(\beta)\right\}
\end{equation}
 at least at one point $\beta$. It is clearly seen that all
recorded states satisfy this lowest-order condition for nonclassicality
with a high significance of $s \geq 10$. Hence, we directly observe
signatures of nonclassicality in the characteristic functions.

We note that Fig.~\ref{fig:CF} reveals that the $P$ functions of
these states are highly singular.  This is due to the fact that condition~(\ref{eq:criterion:CF}) is satisfied for all $\beta$ with large modulus,
indicating that $\Phi(\beta)$ is not integrable. Therefore, we cannot expect $P(\alpha)$
to be a well-behaved function. However, if $\Phi(\beta)$ does not satisfy
Eq.~(\ref{eq:criterion:CF}), one may be able to compute its Fourier transform
and check nonclassicality of the state based on the failure of the
$P$ function to be nonnegative, see \cite{SPATS}.

\subsection{Theoretical generalization of the results}

Whereas in the experiment only states with Gaussian phase noise are investigated, we now prove that phase-diffused squeezed vacuum states always fulfill condition (\ref{eq:criterion:CF}), indifferent of the phase distribution $p(\varphi)$. Although this criterion is not necessarily fulfilled by an arbitrary nonclassical state, it turns out to be sufficient for any state with a characteristic function of the form (\ref{eq:mixed:state:CF}).

First, we note that $\Phi_{\rm sv}(\beta;\varphi)$ is $\pi$-periodic in the angle $\varphi$. Without any loss of generality, we can assume that the function $p(\varphi)$ in Eq.~(\ref{eq:mixed:state:CF}) can be regarded as a probability density over the interval $[0,\pi]$, i.e.~$p(\varphi) \geq 0$ and $\int_0^\pi p(\varphi) d\varphi = 1$. For technical reasons, we may further assume that $p(\varphi)$ is $\pi$-periodic as the characteristic function of squeezed vacuum, such that the domain of integration in Eq.~(\ref{eq:mixed:state:CF}) may be any interval of length $\pi$.

It is clear that the average value of the function $p(\varphi)$ equals to ${\pi}^{-1}$ on any interval of length $\pi$. As an intermediate result, we show that for any positive integer $n$ there exists an interval $[\varphi_0-\tfrac{\pi}{2n},\varphi_0+\tfrac{\pi}{2n}]$ on which the average of $p(\varphi)$ is greater or equal than ${\pi}^{-1}$:
\begin{equation}
 	\forall n \in \mathbb N\quad \exists \varphi_0 \in \mathbb R: \int_{-\frac{\pi}{2n}}^{\frac{\pi}{2n}} p(\varphi+\varphi_0)d \varphi \geq \frac{1}{\pi}\frac{\pi}{n}=\frac{1}{n}.\label{eq:proof:lemma}
\end{equation}
Let us assume that this claim is wrong. Then we can find a certain $n$ such that
\begin{equation}
 	\forall \varphi_0 \in \mathbb R:\qquad\int_{-\frac{\pi}{2n}}^{\frac{\pi}{2n}} p(\varphi+\varphi_0)d \varphi < \frac{1}{n}
\end{equation}
holds true. With this assumption we can decompose the interval $[0,\pi]$ into a set of $n$ successive intervals of the length $\frac{\pi}{n}$ and calculate
\begin{equation}
	\int_{0}^{\pi} p(\varphi) d \varphi= \sum_{j=1}^n\int_{-\frac{\pi}{2n}}^{\frac{\pi}{2n}}p\left(\varphi + (j-\tfrac{1}{2})\tfrac{\pi}{n}\right)d \varphi	<  \sum_{j=1}^n\frac{1}{n} = 1,
\end{equation}
which is in clear contradiction to the normalization condition of $p(\varphi)$. Consequently, our assumption is wrong, and the desired statement is verified.

Now let us look at the characteristic function defined in Eqs.~(\ref{eq:charfunc:squeezed}) and (\ref{eq:mixed:state:CF}). We assume that the $x$-quadrature is the squeezed one, so $V_x < 1 < V_p$. Hence, we can find a positive integer $n$ such that
\begin{equation}
 \forall\varphi \in [-\tfrac{\pi}{2n},\tfrac{\pi}{2n}]:\quad 1-V_x\cos^2(\varphi)- V_p\sin^2(\varphi) > 0
	\label{eq:phasediff:proof:squeeze:ineq}
\end{equation}
holds. Simultaneously, we can find a real number $\varphi_0$ such that the inequality~(\ref{eq:proof:lemma}) is  fulfilled,
and choose $\beta = |\beta| e^{i\varphi_0}$. Making use of the simple relations
\begin{equation}
\Phi_{\rm sv}(\beta;\varphi) = \Phi_{\rm sv}(\beta e^{-i\varphi};0),\quad \Phi_{\rm sv}(\beta;0) = \Phi_{\rm sv}(\beta^* ;0),
\end{equation}
and taking into consideration that the domain of integration in Eq.~(\ref{eq:mixed:state:CF}) may be an arbitrary interval of length $\pi$, we calculate
\begin{eqnarray}
 	\Phi(|\beta|e^{i\varphi_0}) &=& \int_{\varphi_0-\tfrac{\pi}{2}}^{\varphi_0+\tfrac{\pi}{2}} p(\varphi)\Phi_{\rm sv}(|\beta| e^{i\varphi_0}; \varphi)\,d\varphi\nonumber\\
		&=& \int_{\varphi_0-\tfrac{\pi}{2}}^{\varphi_0+\tfrac{\pi}{2}} p(\varphi)\Phi_{\rm sv}(|\beta| e^{i(\varphi_0- \varphi)}; 0)\,d\varphi \nonumber\\
		&=& \int_{-\tfrac{\pi}{2}}^{\tfrac{\pi}{2}} p(\varphi+\varphi_0)\Phi_{\rm sv}(|\beta| e^{i\varphi}; 0)\,d\varphi.
\end{eqnarray}
Since the integrand is nonnegative, a diminution of the domain of integration decreases the value of the integral:
\begin{equation}
 	\left|\Phi(|\beta|e^{i\varphi_0})\right| \geq \int_{-\tfrac{\pi}{2n}}^{\tfrac{\pi}{2n}} p(\varphi+\varphi_0)\Phi_{\rm sv}(|\beta| e^{i\varphi}; 0)\,d\varphi \label{eq:proof:Phi:greater:step:1}.
\end{equation}
Because $n$ is chosen such that Eq.~(\ref{eq:phasediff:proof:squeeze:ineq}) is fulfilled for each $\varphi$ in the domain of integration, there exists an $\epsilon> 0$ such that
\begin{equation}
 	\forall \varphi \in [-\tfrac{\pi}{2n},\tfrac{\pi}{2n}]:\quad\left|\Phi_{\rm sv}(|\beta|e^{i\varphi})\right| \geq e^{\epsilon |\beta|^2/2}.
\end{equation}
Consequently, we have a lower bound for Eq.~(\ref{eq:proof:Phi:greater:step:1}):
\begin{eqnarray}
 	|\Phi(|\beta|e^{i\varphi_0};0)| &\geq& e^{\epsilon |\beta|^2/2}\int_{-\tfrac{\pi}{2n}}^{\tfrac{\pi}{2n}} p(\varphi+\varphi_0)\,d\varphi \label{eq:proof:Phi:greater:step:2}
\end{eqnarray}
Additionally, we have chosen $\varphi_0$ such that Eq.~(\ref{eq:proof:lemma}) is fulfilled. Therefore,
\begin{equation}
 	|\Phi(|\beta|e^{i\varphi_0})| \geq  \frac{e^{\epsilon |\beta|^2/2}}{n}.
\end{equation}
Obviously, the characteristic function $\Phi(\beta)$ is not bounded, independently of the phase noise distribution $p(\varphi)$. Consequently, we are always able to prove nonclassicality by means of Eq.~(\ref{eq:criterion:CF}), which is the most simple criterion among a necessary and sufficient hierarchy \cite{VogelCFHierarchy}.

\section{Nonclassicality in terms of moments}

Besides the signatures of nonclassicality in {terms of} the characteristic function,
different criteria for demonstrating nonclassicality are known. Since
we measured {time series of individual quadrature values we can calculate all
higher-order moments of the quadrature operator $\hat x$.}

\subsection{Hong-Mandel squeezing}

We {begin by examining higher-order squeezing as proposed by Hong and Mandel
\cite{HongMandel}.} To this end, we calculate the degree of $2n$-th order
squeezing
\begin{equation}
        q_{2n} = \frac{\left<(\Delta \hat x)^{2n}\right>}{(2n-1)!!}
        -1,\qquad n \in\mathbb N
\end{equation}
 where $\Delta \hat x = \hat x - \left<\hat x\right>$. The
moments can be estimated from the sample of quadrature data quite
naturally by replacing expectation values by their arithmetic means. It
is sufficient to verify nonclassicality if at least one of the $q_{2n}$
is negative.

\begin{table*}
\footnotesize
\begin{tabular}{c || c | c | c | c | c}
        $\sigma / {}^{\circ}$   &       $q_2$      & $q_4$    &       $q_6$      & $q_8$ & $q_{10}$\\\hline
$0.0$		&	$-0.6362(1\pm0.3\%)$ &	$-0.8667(1\pm0.16\%)$ &	 $-0.9506(1\pm0.12\%)$ &	$-0.9813(1\pm0.09\%)$ &	 $-0.9927(1\pm0.07\%)$ \\
$6.3$		&	$-0.5717(1\pm0.04\%)$ &	$-0.8090(1\pm0.03\%)$ &	 $-0.9102(1\pm0.03\%)$ &	$-0.9549(1\pm0.04\%)$ &	 $-0.9754(1\pm0.06\%)$ \\
$12.6$	&	$-0.4060(1\pm0.08\%)$ &	$-0.5509(1\pm0.15\%)$ &	 $-0.5459(1\pm0.60\%)$ &	$-0.3852(1\pm4.2\%)$ &	 $0.0798(1\pm95\%)$ \\
$22.2$	&	$0.0196(1\pm3.2\%)$ &	$0.6864(1\pm0.53\%)$ &	 $2.982(1\pm0.84\%)$ &	$10.61(1\pm1.7\%)$ &	 $37.27(1\pm3.3\%)$ \\
$\infty$	&	$1.908(1\pm0.09\%)$ &	$10.68(1\pm0.16\%)$ &	 $51.72(1\pm0.32\%)$ &	$249.6(1\pm0.65\%)$ &	 $1222(1\pm1.23\%)$
\end{tabular}
\normalsize
\caption{Degree of squeezing $q_{2n}$ for different orders $2n$ and standard deviations of phase noise $\sigma$.}
\label{tab:q:parameter}
\end{table*}

Table~\ref{tab:q:parameter} shows the degree of squeezing for different
orders and different phase noise {strengths}. Obviously, only if the lowest-order
parameter $q_2$ is negative, then the parameter $q_{2n}$ of higher order
can also be negative. Hence, we may only observe higher-order
squeezing if the state already shows standard squeezing. The investigation
of the degree of higher-order squeezing does not extend the range of
detection of nonclassicality of phase-diffused squeezed states. This
is not surprising, since it can be shown that for Gaussian states
(\ref{eq:Wigner:squeezed}) the degree of squeezing is given by
\begin{equation}
        q_{2n}(\varphi) = \left[V_x \cos(\varphi) + V_p
        \sin(\varphi)\right]^n - 1.
\end{equation}
For these states, squeezing always implies higher-order
squeezing and vice versa \cite{HongMandel}. Phase diffusion can only
smooth out the phase dependence of the moments and diminish the nonclassical effect.

\subsection{Matrices of normally ordered moments}

Normally ordered moments of the quadrature operator $\hat x$ can be
estimated from measured data points $\{x_j\}_{j=1}^N$ via appropriate
sampling relations (see Appendix \ref{app:sampling}),
 \begin{equation}
        \left<:\hat x^k:\right> \approx \frac{1}{2^{k/2}N}\sum_{j=1}^N
        H_k\left(\tfrac{x_j}{\sqrt{2}}\right),
\end{equation}
where $H_k(x)$ are the Hermite polynomials. With these
moments at hand, we can examine the nonclassicality criterion of Agarwal
\cite{AgarwalMatrices}. It has been shown that a state is nonclassical
if at least one of the matrices
 \begin{equation}
  M^{(l)} = \left(\begin{array}{c c c c}
                      1                                         &
                      \left<:\hat x:\right>   &       \ldots  &
                      \left<:\hat x^{l-1}:\right>\\ \left<:\hat x:\right>
                      &       \left<:\hat x^2:\right> &       \ldots  &
                      \left<:\hat x^{l}:\right>\\
                                        \vdots                          &
                                        \vdots                          &
                                        \ddots  &       \vdots\\
                      \left<:\hat x^{l-1}:\right>       &
                      \left<:\hat x^l:\right> &       \ldots  &
                      \left<:\hat x^{2l-2}:\right>\\ \end{array}\right)
\end{equation}
 is not positive semidefinite. This can be
verified by showing that at least one of the principal minors of such
a matrix is negative \cite{ShchukinVogel}. However, to this end we had
to check up to $2^{l} - 1$ principal minors for each matrix $M^{(l)}$,
which is a computationally expensive task.

Here we use the fact that the existence of a negative eigenvalue of $M^{(l)}$ demonstrates the violation of positive semidefiniteness. Therefore, we determine
\begin{equation}
        \lambda_{\rm min}^{(l)} = \min_{\vec x \neq \vec 0}\frac{\vec x^T M^{(l)}\vec x}{\vec x^T\vec x}
\end{equation}
via a conjugate gradient algorithm, see, e.g., \cite{MinEigenvalueAlgorithm}. It can be shown that $\lambda_{\rm min}^{(l)}$ equals to the minimum eigenvalue of $M^{(l)}$. In this way, we only need to calculate one quantity per matrix to examine its definiteness. Its standard deviation is determined by using a bootstrap method: We generate new quadrature data, distributed as the experimentally measured quadratures, $100$ times to obtain a statistical sample of eigenvalues, which gives the standard deviation, cf.~\cite{Bootstrap}.

% \begin{figure}[!ht]
%       \centering
%       \setlength{\subfigcapskip}{4pt}
%       \subfigure[$2\times2$ Matrix]{
%       \includegraphics[width=7cm]{MinEigenvalue2}}
%       \hspace{5pt}
%       \subfigure[$4\times4$ Matrix]{
%       \includegraphics[width=7cm]{MinEigenvalue4}}\\
%
%       \subfigure[$6\times6$ Matrix]{
%       \includegraphics[width=7cm]{MinEigenvalue6}}
%       \hspace{5pt}
%       \subfigure[$8\times8$ Matrix]{
%       \includegraphics[width=7cm]{MinEigenvalue8}}
%       \caption{Minimum eigenvalues of Agarwal's matrices $M^{(l)}$ for different $l$ and phase noise $\sigma$. }
%       \label{fig:MinEigenvalues}
% \end{figure}

\begin{table*}
\footnotesize
\begin{tabular}{c || c | c | c | c | c}
        $\sigma / {}^{\circ}$   &       $2\times 2$ Matrix      & $4\times 4$ Matrix    &       $6\times 6$ Matrix      & $8\times 8$ Matrix & $10\times 10$ Matrix\\\hline
$0.0$           & $-0.6362(1\pm0.25\%)$ &       $-4.294(1\pm0.86\%)$ & $-104.0(1\pm2.5\%)$      & $-6201(1\pm6.1\%)$  & $-722\cdot 10^3(1\pm12\%)$\\
$6.3$   & $-0.5717(1\pm0.03\%)$ &       $-3.337(1\pm0.11\%)$ & $-69.93(1\pm0.35\%)$     & $-3593(1\pm0.98\%)$ & $-335\cdot 10^3(1\pm2.5\%)$\\
$12.6$  & $-0.4060(1\pm0.08\%)$  &      $-2.040(1\pm1.1\%)$      & $-6.728(1\pm53\%)$   & $-107.4(1\pm110\%)$  & $-1259\cdot 10^3(1\pm49\%)$\\
$22.2$  & $0.0197(1\pm3.0\%)$  &        $-0.2323(1\pm1.1\%)$ & $-0.5358(1\pm4.1\%)$     & $-2.299(1\pm71\%)$   &  $-459\cdot 10^3(1\pm40\%)$\\
$\infty$&  $1.0000(1\pm0\%)$              &     $0.7856(1\pm1.2\%)$      & $0.5493(1\pm12\%)$   & $10.85(1\pm13\%)$        & $1113(1\pm94\%)$
\end{tabular}
\normalsize
\caption{Table of minimum eigenvalues of the matrices $M^{(l)}$ for different $l$ and phase noise $\sigma$. The existence of significantly negative values indicate the nonclassicality.} %\marker{In der 3. Zeile sinkt der Fehler, wenn man zur 10x10 Matrix geht. Also kann es sein, dass eine 20x20 Matrix die Nichtklassizit\"at komplett diffundierter Zust\"ande auch finden w\"urde. Wie ist dann unsere Argumentation? Z.B. so?: Das Bilden so gro{\ss}er Matrizen ist rechnerisch aufwendig.}}
\label{tab:Minimum:Eigenvalues}
\end{table*}

The experimental results are shown in Table
\ref{tab:Minimum:Eigenvalues}. We only consider matrices of even
dimension, since we noted that the minimum eigenvalues of $M^{(2n)}$
and $M^{(2n+1)}$ are equal. This may be due to the fact that odd
moments of squeezed vacuum states vanish, giving the matrices a special
structure. We observe that all matrices, which belong to the states
showing squeezing, have significantly negative eigenvalues. Hence,
their nonclassical character can be directly observed in the sign of
the smallest eigenvalue. Furthermore,  for the states with $\sigma \geq
22.2^\circ$
%\jd{Should this not be simply greater than: no, since this part of the discussion covers only the first column of the table (2x2-Matrix)??}
the matrix $M^{(2)}$ is positive semidefinite, since this
directly corresponds to the absence of  quadrature squeezing. However,
for the state with $\sigma = 22.2^\circ$ the matrices $M^{(l)}$ with
$l\geq 4$ possess a negative eigenvalue. Therefore, Agarwal's criterion
extends the range of detection of nonclassicality. Only for the completely
phase-diffused state, are we not able to prove nonclassicality by this
method. For this state, the effect might appear in higher-dimensional matrices, but the statistical uncertainty
might hide the effect.

\section{Conclusion}

We have used experimental data sets of quadrature measurements on phase-diffused squeezed states for a test of different nonclassicality criteria. Even for a \emph{completely} phase-diffused squeezed state, i.e.~where the measured statistics were identical for all homodyne detection phase angles, we found a pronounced nonclassical character. This could be illustrated with the help of the characteristic function of the $P$ function: It directly shows nonclassical features in the lowest-order criterion of \cite{VogelCFHierarchy} and the nonclassicality was detected with a rather high signal-to-noise ratio.  Other nonclassicality criteria, such as higher-order squeezing or the violation of positive semidefiniteness of Agarwal's matrices, fail to reveal nonclassicality beyond squeezing or only show nonclassical behavior in matrices of higher dimension. Therefore, we demonstrated for the radiation under study,
that the characteristic function of the $P$ function, which contains
information about all moments of the state, can be a more powerful tool
for the examination of nonclassicality than a finite set of moments.

Eventually, we note that the evaluation of the statistical significance of nonclassical effects is much easier in terms of characteristic functions, since Eq.~(\ref{eq:Sample:CF:Standard:Deviation}) provides a simple relation between the variance and the value of $\Phi(\beta)$. Testing the definiteness of matrices of moments requires complex nonlinear procedures, for instance the calculation of the smallest eigenvalue or the principal minors. This leads to complications in the estimation of the statistical significance, which increase the computational effort. For matrices of high orders the resulting errors are large and they may hide the sought nonclassical effects.

% \marker{K\"onnen wir eine Aussage machen, ob in unserem Purifikations-Experiment nach der charakteristischen Funktion die Nichtklassizit\"at erh\"oht wurde? Oder ist die charakteristische Funktion gar kein monotones Mass?}
% \marker{ODER BRAUCHEN WIR DAF\"UR WEITERE UNTERSUCHUNGEN UND EIN WEITERES PAPER?}

\section*{Acknowledgements}

We thank J.~Fiur\'{a}\v{s}ek for many helpful discussions.

\appendix

\section{Sampling formula for normally ordered moments}

\label{app:sampling}

The characteristic function $\Phi(\beta)$ can be given in terms of normally ordered moments of the creation and annihilation operator \cite{vowe-book}:
\begin{eqnarray}
        \Phi(\beta)%&=& \sum_{k= 0}^\infty \frac{1}{k!}\sum_{l=0}^k\binom{k}{l}\beta^l(-\beta^*)^{k-l} \left<\hat a^{\dagger l}\hat a^{k-l}\right>\nonumber\\
                &=& \sum_{k= 0}^\infty \frac{1}{k!} \left<:\left(\beta\hat a^{\dagger} -\beta^* \hat a\right)^k:\right>.
\end{eqnarray}
Introducing the phase-dependent quadrature operator $\hat x(\varphi) = \hat a^\dagger e^{-i\varphi} + \hat a e^{i\varphi}$, we have
\begin{equation}
        \Phi(\beta)= \sum_{k= 0}^\infty \frac{(i|\beta|)^k}{k!} \left<:\hat x(\tfrac{\pi}{2}-\arg(\beta))^k:\right>.
\end{equation}
Consequently, the normally ordered moments $\left<:\hat x(\varphi)^k:\right>$ can be calculated from the characteristic function of the $P$ function as
\begin{equation}
        \left<:\hat x(\varphi)^k:\right> = \frac{\partial^k}{\partial (ib)^k} \Phi(i b e^{-i\varphi})\bigg|_{b = 0}.
                \label{eq:calc:moments:from:Phi}
\end{equation}
To obtain a formula which can be applied in practice, we insert Eq.~(\ref{eq:charfunc:P}) into Eq.~(\ref{eq:calc:moments:from:Phi}) and use the definition of the Hermite polynomials in the form  $(-1)^k H_k(\xi)e^{-\xi^2} = \frac{\partial^k}{\partial \xi^k}e^{-\xi^2}$ with $\xi = ib/\sqrt{2}$.  Neglecting the phase argument, we find
\begin{eqnarray}
        \left<:\hat x^k:\right> &=&  \frac{\partial^k}{\partial (ib)^k} \left<e^{ib\hat x}\right> e^{b^2/2}\bigg|_{b = 0}\nonumber\\
  &=& \left<\frac{\partial^k}{\partial (\sqrt{2}\xi)^k} e^{-\left[\xi - \hat x/\sqrt{2}\right]^2} e^{\hat x^2/2}\right>\bigg|_{\xi = 0}\nonumber\\
  &=& \left<\frac{(-1)^k}{2^{k/2}}H_k\left(\xi - \tfrac{\hat x}{\sqrt{2}}\right)e^{-\left[\xi - \tfrac{\hat x}{\sqrt{2}}\right]^2} e^{\tfrac{\hat x^2}{2}}\right>\bigg|_{\xi = 0}\nonumber\\
 %% &=& \left<\frac{(-1)^k}{2^{k/2}}H_k\left(-\tfrac{\hat x}{\sqrt{2}}\right)e^{-\hat x^2/2} e^{\hat x^2/2}\right>\nonumber\\
&=& \frac{1}{2^{k/2}}\left<H_k\left(\tfrac{\hat x}{\sqrt{2}}\right)\right>.
\end{eqnarray}
Hence, we obtain normally ordered moments from measured quadratures via
\begin{equation}
        \left<:\hat x^k:\right> \approx \frac{1}{2^{k/2}N}\sum_{j=1}^N H_k\left(\tfrac{x_j}{\sqrt{2}}\right),
\end{equation}
The approximation sign indicates that the right hand side is a statistical estimator.

\end{document}